\documentclass[twocolumn,twoside,slac_two]{revtex4}
\usepackage{graphicx}
\usepackage{fancyhdr}
\newcommand{\neut}{\mbox{$\tilde{\chi}_1^0$}}
\newcommand{\pt}{\mbox{$p_T$}}

\begin{document}

\title{Searches for Charginos and Neutralinos with the D0 Detector}

%

\author{T. Adams \em for the D0 Collaboration}
\affiliation{Department of Physics, Florida State University, Tallahassee, FL 32306, USA}

\begin{abstract}
Within the framework of supersymmetry, charginos and/or neutralinos are
often the preferred topics of searches for experimental evidence.  This
is due to the facts that in much of parameter space they are the 
lightest supersymmetric partners and they offer unique final states
to separate from standard model backgrounds.  The D0 experiment has
performed several recent searches including the traditional trilepton
final state and a decay chain involving dark photons.
\end{abstract}

\maketitle

\thispagestyle{fancy}


\section{Introduction}
The beyond the standard model (BSM) of supersymmetry (SUSY) has
been explored theoretically~\cite{bib:susyprimer} for several 
decades.  While no experimental evidence has been found, searches
for supersymmetry continue to be an important part of the 
physics programs at high energy colliders.

The spectrum of sparticle masses is determined by the SUSY
model and the choice of parameters.  However, in many variations
(e.g. mSUGRA) the lightest sparticles may be charginos and 
neutralinos, mixtures of the wino and bino which are the
superpartners of the $W$ and $Z$ bosons.  If the lightest
neutralino \neut\ is also the lightest
supersymmetric particle (LSP) and $R$-parity is conserved,
then the \neut\ escapes detection in collider experiments
and appears only as a contribution to missing transverse
energy.  Therefore, pair production of \neut\ becomes
difficult to observe.  A more interesting process is associated
production of a chargino and the second lightest
neutralino ($p\bar{p} \to \tilde{\chi}_1^\pm \tilde{\chi}_2^0$)
where some of the decay products of the $\tilde{\chi}_1^\pm$ 
and $\tilde{\chi}_2^0$ can be observed.

The D0 experiment at the Fermilab Tevatron has collected data
on $p\bar{p}$ collisions at $\sqrt{s}=1.96$ TeV since 2001.  The
detector consists of an inner tracking system (with solenoid
magnet), a liquid argon calorimeter, and an outer muon 
spectrometer.  A full description is available in
Ref.~\cite{bib:d0detector}.  

\section{Trileptons}

\begin{figure*}
\centering
\includegraphics[width=135mm]{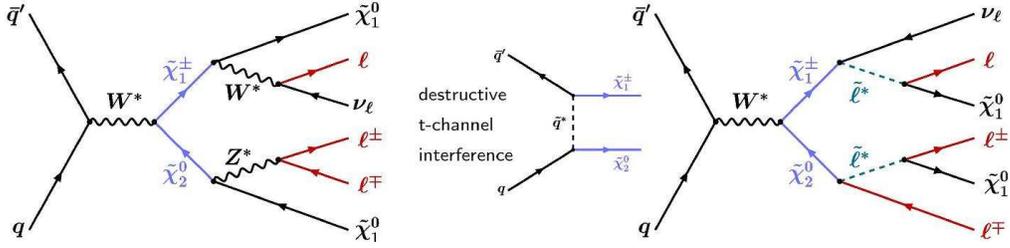}
\caption{Feynman diagrams for the associated production of a chargino
and a neutralino with decay into the trilepton final state.} 
\label{fig:trilepfeyn}
\end{figure*}

A traditional search for associated chargino and neutralino
production involves looking for events with three leptons
plus missing transverse energy~\cite{bib:trilepton}.  
In this case, the chargino
decays via $\tilde{\chi_1}^\pm \to \ell^\pm \nu \tilde{\chi}_1^0$ 
while the neutralino decays via
$\tilde{\chi}_2^0 \to \ell^{\prime\pm} \ell^{\prime\mp} \tilde{\chi}_1^0$
(Fig.~\ref{fig:trilepfeyn}).
Here, $\ell$ and $\ell^\prime$ may or may not be the same
lepton flavor while the neutrinos and lightest neutralinos escape
undetected.

\begin{figure}
\centering
\includegraphics[width=80mm]{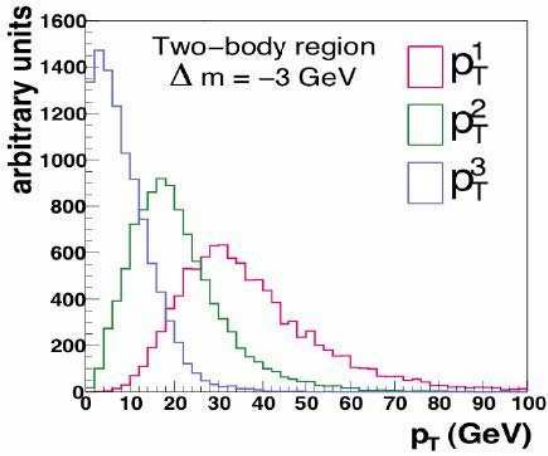}
\caption{Distribution of charged lepton \pt\ for the trilepton
final state.  The red histogram is the highest \pt\ lepton, the
green the second highest \pt\ lepton, and the blue is the third
highest \pt\ lepton.} \label{fig:trilep_pt}
\end{figure}

Six final state particles means that some of the time,
one or more of the charged leptons has low \pt.  This can be particularly
true depending on the sparticle mass relationships.  
Figure~\ref{fig:trilep_pt} shows the \pt\ of the three charged
leptons (in order of decreasing \pt) for one model point.  Therefore,
techniques have been developed to include low \pt\ leptons in
the analysis.  In this case, we allow the third lepton to be
identified as an isolated track with low \pt.  This recovers
part of the acceptance that would be lost by requiring it to
be identified as a high quality lepton.  Another technique (not
used here) is to search for like sign leptons.

One reason this channel is considered a ``golden channel'' is
that there are very few standard model processes that may
produce trilepton events.  The primary source of events with
three real leptons is via diboson production (e.g. $WZ$) which
has a small cross section.  Single boson production ($W$ or
$Z$) contributes only through a fake third isolated track
and/or fake missing transverse energy.

\subsection{Event selection}

Event selection criteria are optimized in two regions,
one ``low-\pt" and one ``high-\pt", to take advantage of the
different kinematics in different regions of mSUGRA
parameter space.  The data is searched in four final states:
(1) di-electron plus lepton ($ee\ell$), (2) di-muon plus
lepton ($\mu\mu\ell$), (3) electron plus muon plus lepton
($e\mu\ell$), and (4) muon plus tau ($\mu\tau$).  The third
lepton is identified as an isolated track without using the
lepton identification criteria.  Details on the selection
criteria are given in Ref.~\cite{bib:trilepton}.

The inclusion of the $\mu\tau$ channel is new for the trilepton
analysis.  Here, hadronic decays of the tau are considered.  
Leptonic decays and single hadron decays ($\tau\to\pi^\pm\nu$)
had previously been included when they passed
other lepton requirements.  The inclusion of this final state
allows for greater sensitivity at higher values
of $\tan\beta$.  The $\mu\tau$ channel is broken down into
two subsets: $\mu\tau\tau$ and $\mu\tau\ell$ where the difference
is whether a second reconstructed tau or a isolated track
is required.  The $\mu\tau$ channels are not optimized
separately for low and high \pt.

\subsection{Results}

\begin{figure}
\centering
\includegraphics[width=80mm]{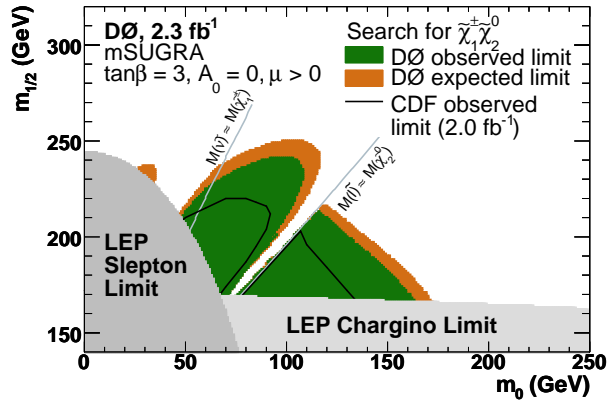}
\caption{95\% C.L. limits on the mSUGRA model in the m$_{1/2}$ versus
m$_0$ plane.  The orange indicates the expected limits while the
green shows the observed limits.  Previously published results
from LEP and CDF are also shown.} \label{fig:trilepton_plane}
\end{figure}

\begin{figure}
\centering
\includegraphics[width=80mm]{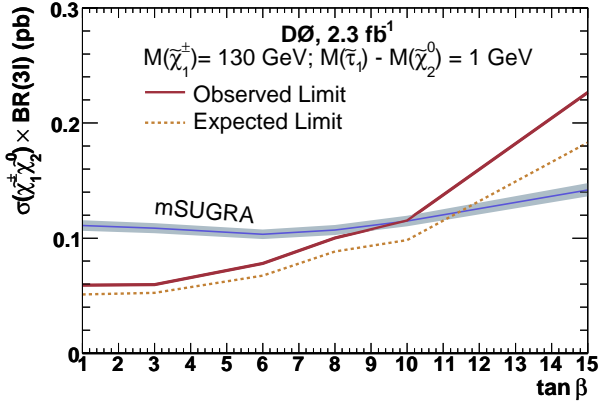}
\caption{95\% C.L. limits on the cross section times branching
ratio as a function of $\tan\beta$ for a chargino mass of
130 GeV and $m(\tilde{\tau})-m(\tilde{\chi}^0_2 = 1$ GeV.  The
cross section for the mSUGRA model is shown as the blue line.} 
\label{fig:trilepton_beta}
\end{figure}

Comparisons of data and expected background for the various channels
is given in Tab.~\ref{tab:trilepton_results}.  Good agreement is
observed.  From this we set limits on the parameters of the 
mSUGRA model.  Figure~\ref{fig:trilepton_plane} shows the limits
in the m$_{1/2}$ versus m$_0$ plane.  Figure~\ref{fig:trilepton_beta}
shows the limit on the cross section times branching ratio as a
function of $\tan\beta$.

\begin{table}[h]
\begin{center}
\caption{Numbers of events for data and expected background for the 
four final states and both
the low-\pt\ and high-\pt\ optimizations in the trilepton search.}
\begin{tabular}{|l|c|c|} \hline 
& & \textbf{Expected} \\ 
& ~~~\textbf{Data}~~~ & \textbf{~~~Background~~~} \\ \hline 
$ee\ell$ & &  \\
~~low-\pt  & 2 & 1.8 $\pm$ 0.2 \\
~~high-\pt~~~~ & 0 & ~~~0.8 $\pm$ 0.1~~~ \\ \hline
$e\mu\ell$ & &  \\
~~low-\pt  & 2 & 0.8 $\pm$ 0.2 \\
~~high-\pt & 0 & 0.5 $\pm$ 0.1 \\ \hline
$\mu\mu\ell$ & &  \\
~~low-\pt  & 4 & 1.2 $\pm$ 0.2 \\
~~high-\pt & 4 & 2.0 $\pm$ 0.3 \\ \hline
$\mu\tau$ &  & \\
~~$\mu\tau\tau$ & 1 & 0.8 $\pm$ 0.2 \\
~~$\mu\tau\ell$ & 0 & 0.8 $\pm$ 0.1 \\ \hline
\end{tabular}
\label{tab:trilepton_results}
\end{center}
\end{table}

\section{Dark photons}

Recent experimental evidence for an excess of positrons and/or
electrons within the cosmic ray spectrum~\cite{bib:pamela,bib:atic}
have inspired a new model of a ~1 TeV dark matter candidate that
can annihilate with itself.  This annihilation can create two light ($<$ 3 GeV)
gauge bosons called dark photons that are force carriers of a hidden valley
sector~\cite{ArkaniHamed:2008qn,bib:strassler}.  
These gauge bosons can decay via mixing 
with the standard model photon to produce pairs of standard model
fermions.  The branching ratios into fermion types depends upon
the mass of the dark photon.  

\begin{figure}
\centering
\includegraphics[width=80mm]{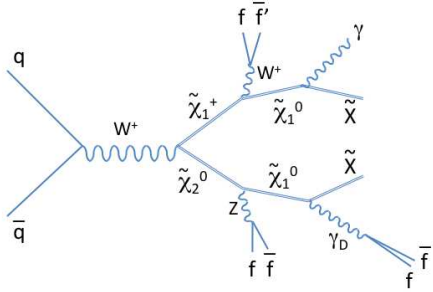}
\caption{Feynman diagram for the associated production of a chargino 
and a neutralino with decays into a dark sector.} \label{fig:darkphoton_feyn}
\end{figure}

D0 has searched for evidence of dark photons through pairs of
electrons or muons with very small opening angle
using 4.1 fb$^{-1}$ of data~\cite{bib:d0darkphoton} .
Figure~\ref{fig:darkphoton_feyn} shows a Feynman diagram 
for the production and decay of dark photons at the Tevatron.
In this scenario, the production still creates an associated
chargino and neutralino (as in the previous search), but
the observable final state is significantly different.

\subsection{Selection criteria}

Events are selected by requiring a photon with \pt\ $>$ 30 GeV and
missing transverse energy $>$ 20 GeV.  Pairs of oppositely signed
tracks with ${\cal R} < 0.2$ (where ${\cal R} = \sqrt{(\Delta\eta)^2 + 
(\Delta\phi)^2}$) and $\Delta z_{vertex} < 2$ cm are dark photon 
candidates.  They must have momentum greater than 10(5) GeV
for the leading(2nd leading) track.  
Previous analyses likely would have missed such a signal
due to isolation criteria.  Here, isolation variables are
calculated after accounting for the second nearby particle.

Dark photon candidates are divided into two types: electron or
muon.  In the electron case, the tracks must match a single EM
cluster (since the two electrons overlap in the calorimeter).  
In the muon case, one of the tracks must be matched to a 
reconstructed muon.

\subsection{Results}

\begin{figure}
\centering
\includegraphics[width=80mm]{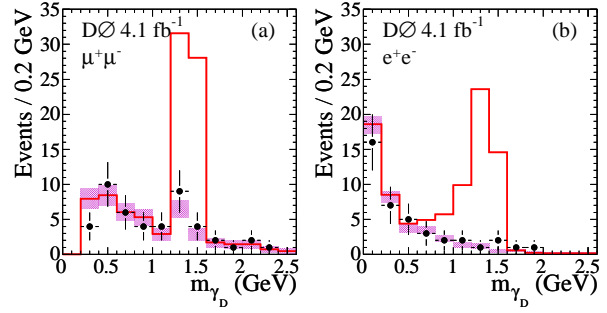}
\caption{Distribution of the invariant mass of dark photon
candidates.  The left distribution shows the dimuon spectrum 
while the right shows the dielectron spectrum.
The background (filled band) is estimated by 
combining three samples with inverted selection criteria.  An
example signal ($m_{\gamma_D} = 1.4$ GeV) added to the background
is shown as the open histogram.} \label{fig:darkphoton_mass}
\end{figure}

The invariant mass distribution for the electron or muon pairs
is studied for evidence of a dark photon resonance.  The
background estimate is created by combining three data samples
with one or more selection criteria inverted.  
Figure~\ref{fig:darkphoton_mass} shows the data, background
estimate, and simulated signal.  No evidence for a narrow
resonance is seen.

\begin{figure}
\centering
\includegraphics[width=80mm]{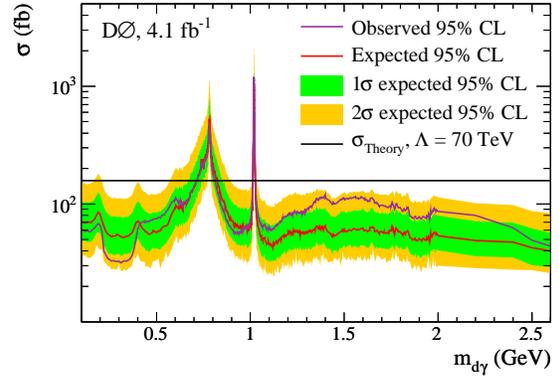}
\caption{Limit on the cross section for dark photon production cross
section as a function of dark photon mass.} 
\label{fig:darkphoton_limit0}
\end{figure}

\begin{figure}
\centering
\includegraphics[width=80mm]{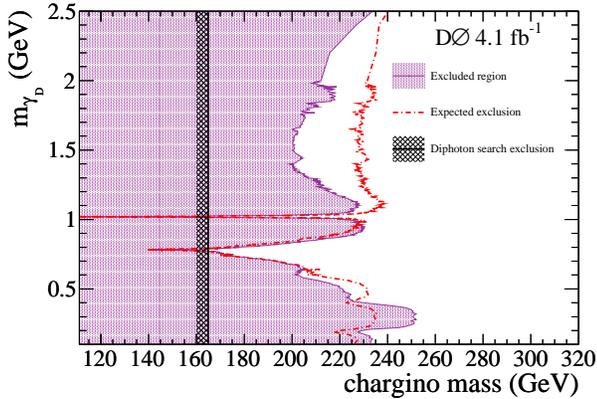}
\caption{Limit on the dark photon mass versus chargino mass for a 
neutralino to dark photon branching ratio of 0.5.  The expected limit
is shown by the dash-dotted line.} \label{fig:darkphoton_limit1}
\end{figure}

\begin{figure}
\centering
\includegraphics[width=80mm]{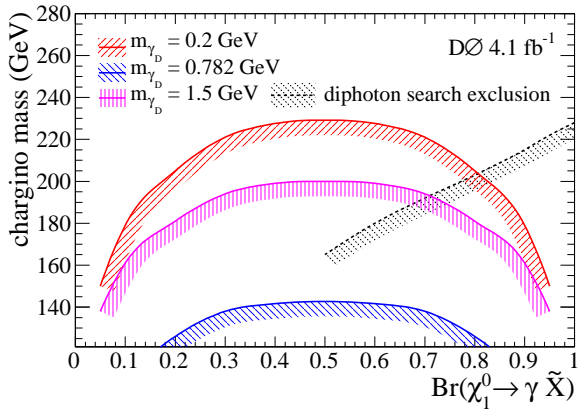}
\caption{Limit on the chargino mass as a function of the neutralino
to dark photon branching ratio for three different dark photon masses.
The limit from a previous diphoton search~\cite{bib:d0diphoton} is
shown as the black contour.} \label{fig:darkphoton_limit2}
\end{figure}

Limits on the production cross section are extracted from the 
invariant mass distributions (Fig.~\ref{fig:darkphoton_limit0}.  
Figure~\ref{fig:darkphoton_limit1}
shows the limits on the dark photon mass as a function of the
chargino mass for a branching ratio of $\tilde{\chi}_1^0 \to 
\gamma_D \tilde{X}$ of 0.5.  Figure~\ref{fig:darkphoton_limit2}
shows the limit on the chargino mass as a function of this
branching ratio for three different dark photon masses.

\section{Summary}

The D0 experiment has recently completed two searches for 
production of charginos and neutralinos in the Tevatron Run II
data set.  The first was a traditional search in the trilepton
final state including $ee\ell$, $e\mu\ell$, $\mu\mu\ell$ and
$\mu\tau$ channels.  The second was a novel search for spatially
close lepton pairs as a signature of a dark photon resulting from
a neutralino decay.  Neither search observed an excess of data
over expectation and limits on the production cross section and
model parameters were set.





\bigskip 

\end{document}